\documentclass[12pt,preprint]{aastex}

\shorttitle{First detection of doubly deuterated hydrogen sulfide}
\shortauthors{Vastel et al.}

\begin{document}

\title{First detection of doubly deuterated hydrogen sulfide\footnote{This research has been 
supported by NSF grant AST-9980846 to the CSO.}}

\author{C. Vastel\footnote{California Institute of Technology, Downs Laboratory of Physics, MS 320-47, 
   1200 East California Boulevard, Pasadena, CA 91125, US, vastel@submm.caltech.edu}, T.G. Phillips$^2$, 
C. Ceccarelli\footnote{Laboratoire d'astrophysique, Observatoire de Grenoble, BP 53, F-38041, 
Grenoble, France}, and J. Pearson\footnote{Jet Propulsion Laboratory, California Institute of Technology, 
CA 91109-8099, Pasadena, US} }

\begin{abstract}
This work was carried out with using the Caltech Submillimeter Observatory and presents the 
observational study of HDS and D$_2$S towards a sample of Class 0 sources, and dense cores. 
We report the first detection of doubly deuterated hydrogen sulfide (D$_2$S) 
in two dense cores and analyze the chemistry of these molecules aiming to help understand the 
deuteration processes in the interstellar medium. The observed values of the D$_2$S/HDS ratio, 
and upper limits, require an atomic D/H ratio in the accreting gas of 0.1~-~1. The study 
presented in this Letter supports the hypothesis that formaldehyde, methanol and hydrogen 
sulfide are formed on the grain surfaces, during the cold pre-stellar core phase, where the 
CO depleted gas has large atomic D/H ratios. The high values for the D/H ratios are consistent 
with the predictions of a recent gas-phase chemical model that includes H$_3^+$ and its 
deuterated isotopomers, H$_2$D$^+$, D$_2$H$^+$ and D$_3^+$ \citep{roberts03}. 
\end{abstract}


\section{Introduction}

Deuterium-bearing molecules have become the target of many observations in recent years and several 
models have been developed to account for them \citep[e.g.][]{tielens83,roberts00a,roberts00b}. 
Twenty-six such molecules have been detected to date in interstellar clouds. The interest 
in this topic lies in the unusual chemistry at work in the cold regions where CO is apparently strongly 
depleted \citep{bacmann02,bacmann03} and in the general question of how much deuterium is trapped 
in the cold, dense phase of the ISM \citep{phillips02}.\\
The study of doubly deuterated interstellar molecules has been very active in the last few
years, since the surprising discovery of a large amount ($\sim$ 5\%) of doubly deuterated
formaldehyde (D$_2$CO) in the low mass protostar IRAS 16293-2422 \citep{ceccarelli98,loinard00}. 
This is more than one order of magnitude higher than in Orion KL, 
where D$_2$CO was first detected by \citet{turner90}. This first discovery was
followed by many other studies which confirmed the presence of large amounts of
D$_2$CO \citep[e.g.][]{ceccarelli02} as well as doubly deuterated ammonia (ND$_2$H) 
\citep[e.g.][]{roueff00,loinard01}.
Gas phase chemical models account relatively well for the observations of
the abundances of singly deuterated molecules, but may not be able to completely
reproduce the large deuterations observed for multiply deuterated
molecules. The large deuterations could also be a
product of active chemistry on the grain surfaces as predicted by
\cite{tielens83}: deuteration during the mantle formation phase 
followed by evaporation of the mantles ices resulting from the heating of
the newly formed star, with injection into the gas phase of the deuterated species 
\citep[e.g.][]{ceccarelli01}.\\
Recently, the detection of a triply deuterated molecule (ND$_3$) extended the observational 
framework. Until now, the possibility for detecting triply deuterated
molecules seemed so remote that their lines were omitted in the spectroscopic catalogs for
astrophysics. The ground-state rotational transition at 309.91 GHz of ND$_3$ has been
detected with the CSO towards the Barnard 1 cloud \citep{lis02}
and the NGC 1333 IRAS 4A region \citep{vdt02}, with
abundance ratios [ND$_3$]/[NH$_3$] $\sim$ 10$^{-3}$ and [ND$_3$/H$_2$] $\sim$ 10$^{-11}$. 
The process at work, chemical fractionation, takes place in cold gas where the deuterium 
substituted molecule is favored due to the lower vibrational uncertainty energy of the heavier 
molecule. The chemical process requires the presence of H$_3^+$ which is usually limited by 
reactions with molecules such as CO. However, in these cold ($\sim$ 10 K) regions of the 
interstellar medium, molecules such as CO accrete onto grain surfaces to form mantles, so 
allowing the gas phase process of deuterium fractionation to take place \citep{bacmann03}. 
In time, the material of the mantle will suffer its own low temperature deuterium fractionation 
\citep{tielens83,rodgers02}. Thus the grains may be important, both in enabling the gas phase 
process through depletion of reactive molecules and also by their own surface chemistry. For the 
latter to be effective, the deuterated molecules on the grain surface must be released, by some 
external energetic process, back into the gas phase, when they can be detected. 
The gas phase mechanism has recently received a boost from the recognition that inclusion of 
the more highly deuterated species D$_2$H$^+$ and D$_3^+$ \citep{phillips02} in the chemical 
calculations, increases at least by a factor of 2 the gas phase D/H ratio \citep{roberts03}.\\
The identified most abundant components of the molecular mantles found on interstellar dust 
grains are H$_2$O and CO \citep[for references, see][]{teixeira99}. The large range of 
condensation temperatures \citep[see][]{nakagawa80} spanned by H$_2$O (T$_c$ $\sim$ 90~K) and 
CO (T$_c$ $\sim$ 16~K) suggests that molecules with intermediate condensation temperatures could 
also be present in the grain mantles. For example, H$_2$S has a condensation temperature of 
$\sim$ 39~K \citep{nakagawa80}. This molecule and other sulfur-bearing molecules have been observed 
in the gas phase in cold, dark clouds where high-temperature processes should not be important 
\citep{minh89,irvine89,fuente90}: x(H$_2$S) = 3 $\times$ 10$^{-9}$ at the SO peak of L134N, 
$<$ 5 $\times$ 10$^{-10}$ at the cyanopolyyne peak towards TMC1 and 
$\sim$ 7 $\times$ 10$^{-10}$ towards the NH$_3$ peak position of TMC1. These values can 
be compared to the range 10$^{-11}$~-~10$^{-10}$ obtained for steady-state gas-phase models 
\citep{millar90,millar97}. Therefore, it poses a problem for gas phase chemistry. The gas-phase 
reactions proposed for the formation of H$_2$S are endothermic (see Section 3) and other mechanisms 
such as shock chemistry or high temperature chemistry are unlikely to be important in dark clouds. 
Grain surface reactions have then been suggested as an alternative pathway to H$_2$S molecule 
formation \citep{duley80}. \\
Because sulfur is chemically similar to oxygen, it is tempting to study this molecule to gain 
insight to the H$_2$O deuteration processes. Like H$_2$O, it is possible that H$_2$S originates 
on grain surfaces, where it was formed by accretion and hydrogenation of sulfur atoms during core 
formation \citep{tielens87}.

\section{Observations and results}

\begin{figure}
\plotone{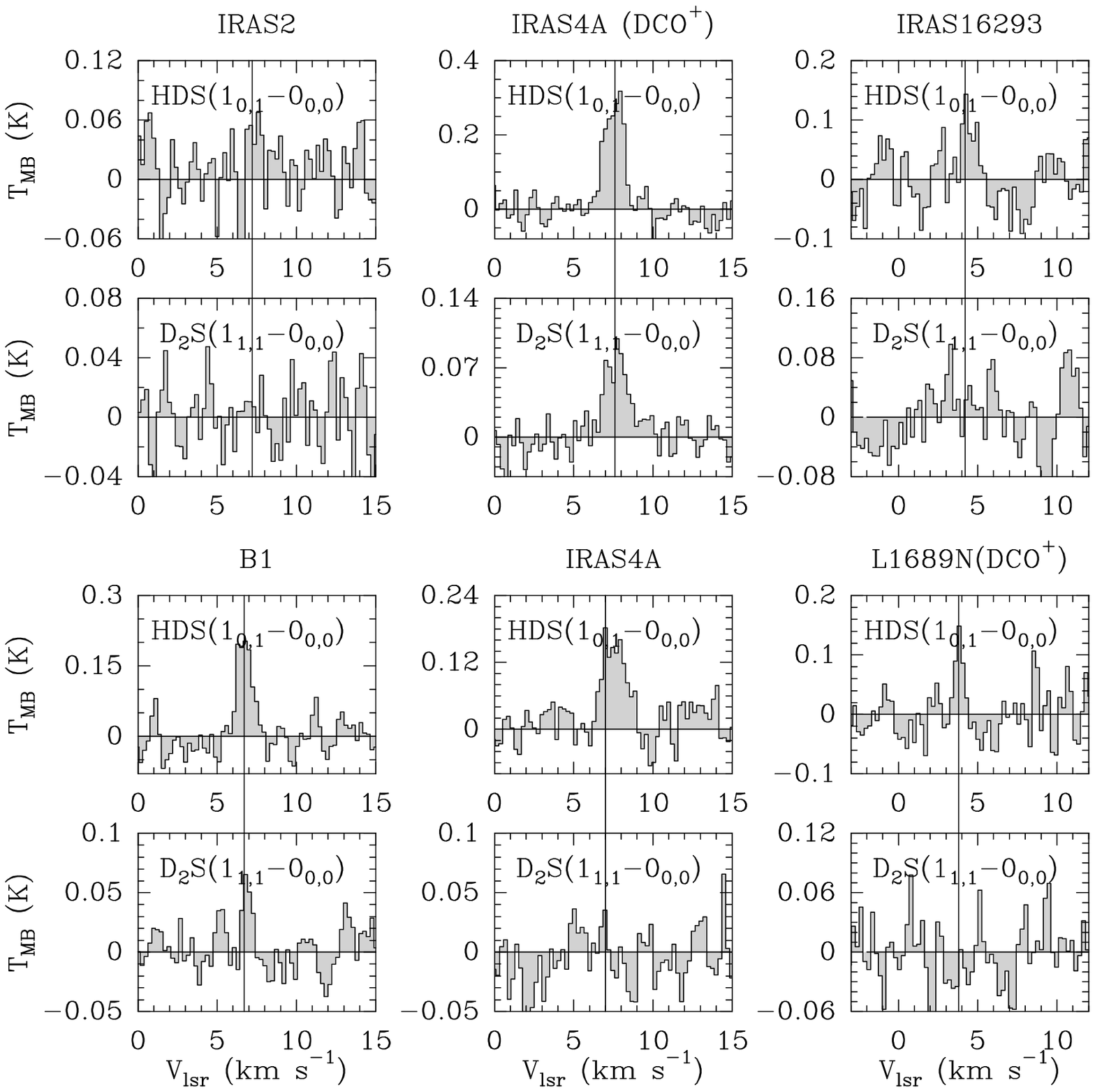}
\caption{Spectra acquired with CSO of transitions of HDS and D$_2$S towards a sample 
of class 0 and dense cores. X-axis represents the velocity in the local standard of rest and 
Y-axis represents the main beam temperature in Kelvins.\label{fig1}}
\end{figure}

The comparison between H$_2$S, HDS and D$_2$S would be useful for a complete study of the 
deuteration processes. In this letter, we report the observations of the HDS and D$_2$S 
ground transitions, accessible at the Caltech Submillimeter Observatory (CSO). 
This observatory is a 10.4 meter single-dish antenna located on top of 
Mauna Kea in Hawaii. At 240 GHz, the telescope has an FWHM beam size of 31$^{\prime\prime}$. 
The observations of these molecules were carried out between March 2002 and September 2002. 
Weather conditions were average with zenith opacities between 0.05 and 0.2 at 225 GHz. \\
The frequency of the observed lines is 244555.58~MHz for 
HDS (1$_{0,1}$~-~0$_{0,0}$) with E$_u$ = 11.7~K, and 237903.6752~MHz for the ortho line of 
D$_2$S (1$_{1,1}$~-~0$_{0,0}$) with E$_u$ = 11.4~K. These are the lowest lying transitions for the two 
molecules. The sample of sources observed includes three class 0 (NGC1333-IRAS2; NGC1333-IRAS4A; 
IRAS16293-2422) and three dense cores (Barnard 1; the peak of emission in DCO$^+$ 3~$\rightarrow$~2 
in the star-forming molecular cloud L1689N, Lis et al., 2002; the peak of emission in 
DCO$^+$ 3~$\rightarrow$~2 near the young stellar object IRAS4A in the NGC1333 complex, 
Lis et al. {\it in preparation}). \\ 
The observation towards IRAS16293-2422 was made in position switching mode with the reference 
position located at $\delta\alpha$ = -180$\arcsec$ and $\Delta\delta$ = 0$\arcsec$ from the 
nominal center whose coordinates are quoted in Table \ref{table1}. 
Positions 180$\arcsec$ or 240$\arcsec$ were observed using the chopping secondary mirror for the 
other sources. The main beam efficiency at 240~GHz is equal to 70$\%$. The spectra obtained towards 
these sources are presented in Figure \ref{fig1} and the measured parameters in Table \ref{table2}.\\
The upper limits on the lines, when not detected, are derived following the relation:

\begin{equation}
3~\sigma~(K) = 3 \times RMS~(K) \times \sqrt{\frac{2 \times binsize~(km~s^{-1})}{\Delta v~(km~s^{-1})}}
\end{equation}
and are reported in Table \ref{table2}. 
From the observed line strengths given in Table \ref{table2}, we can estimate the ratios 
between the D$_2$S and HDS column densities in the observed sources. 
We estimated the optical depth from the observed line intensity using the formula: 
$T_{mb} = [J_{\nu}(T_{ex})-J_{\nu}(T_{bg})](1-e^{-\tau})$
where $J_{\nu}(T) = (h\nu/k)/(e^{h\nu/kT}-1)$ is the radiation temperature of a blackbody at a 
temperature T, and T$_{bg}$ is the cosmic background temperature of 2.7 K. 
We found that the lines are optically thin, even at a low temperature of 10 K. The formula which 
connects the integrated intensity of a rotational transition with the number of emitting molecules 
is, for an optically thin transition and neglecting the background radiation:

\begin{equation}
\int_{}^{}T_{mb}\, dv = \frac{A_{ul}hc^3}{8\pi k\nu^2}N_u
\end{equation}
and, assuming the system is in LTE:
\begin{equation}
N_u=\frac{g_u}{Q(T_{ex})}exp\left(-\frac{E_u}{kT_{ex}}\right)N_{tot}
\end{equation}

The Einstein A coefficient is 1.3~$\times$~10$^{-5}$ s$^{-1}$ for HDS (1$_{0,1}$~-~0$_{0,0}$) 
(JPL catalog) and we computed a value of 5.0~$\times$~10$^{-5}$ s$^{-1}$ for 
D$_2$S (1$_{1,1}$~-~0$_{0,0}$).  
Q(T$_{ex}$) is the partition function $\sum_{i}^{}g_{i}e^{E_{i}/kT_{ex}}$ equal to 
3.1 for HDS (1$_{0,1}$~-~0$_{0,0}$), and 21.6 for D$_2$S (1$_{1,1}$~-~0$_{0,0}$) 
using an excitation temperature of 10 K. This temperature, taken as a kinetic temperature 
is that at which chemical fractionation should be efficient. 
The observed transitions are obviously not enough to calculate the excitation temperature 
in each source. Nevertheless, it is possible to estimate a reliable value for the excitation 
temperature using the kinetic temperature of the gas found in the literature. 
\citet{menten87} \citep[respectively][]{bachiller90} derived a kinetic temperature in L1689N 
(respectively B1) of $\sim$ 12 K based on their ammonia observations. \citet{blake95} estimated 
T$_k$ between 20 and 40 K for the core of IRAS4A based on their CS and H$_2$CO observations. 
\citet{ward96} found that T$_k$ lies in the range of 15-20 K for IRAS2 from their infall model. 
\citet{vandishoeck95} estimated a rotational temperature of $\sim$ 12 K from two HDS lines 
towards IRAS16293. In the three last sources, the Class 0 sources, temperature gradients are 
known to be present, and the quoted estimates refer to the outer regions of the envelopes of 
these sources. Very likely the present HDS and D$_2$S observations probe in fact those external 
regions too. In the following, we will assume that T$_{ex}$ is the same for HDS and D$_2$S, 
namely that the 2 molecules originate in the same region. Luckily enough, the 
calculation of the N(D$_2$S)/N(HDS) ratio does not depend so much on the excitation 
temperature in the 10~-~50~K range, as the energy levels compared are similar.\\
D$_2$S has para and ortho forms. The observed 1$_{1,1}$~-~0$_{0,0}$ transition is the
lowest line in the ortho ladder. To derive the D$_2$S/HDS ratio, the D$_2$S column density 
was calculated for the ortho levels only, and then extrapolated to a total column density 
assuming an ortho to para ratio of 2. The derived D$_2$S/HDS ratios have been computed 
using equations (1~-~3) and are reported in Table \ref{table3}.\\

\section{Chemistry of D$_2$S}

Hydrogen sulfide is traditionally believed to form on the grain
surfaces, since gas phase reactions in cold gas are not
efficient enough to account for the observed H$_2$S abundances
\citep[e.g.][]{tielens87,minh89}.
In dense clouds sulfur is believed to be mostly neutral and initiates the chain reaction:

\begin{equation}
S \stackrel{H_3^+}{\longrightarrow} HS^+ \stackrel{H_2}{\longrightarrow} H_3S^+ \stackrel{e^-}{\longrightarrow} H_2S
\end{equation}

\begin{figure}
\plotone{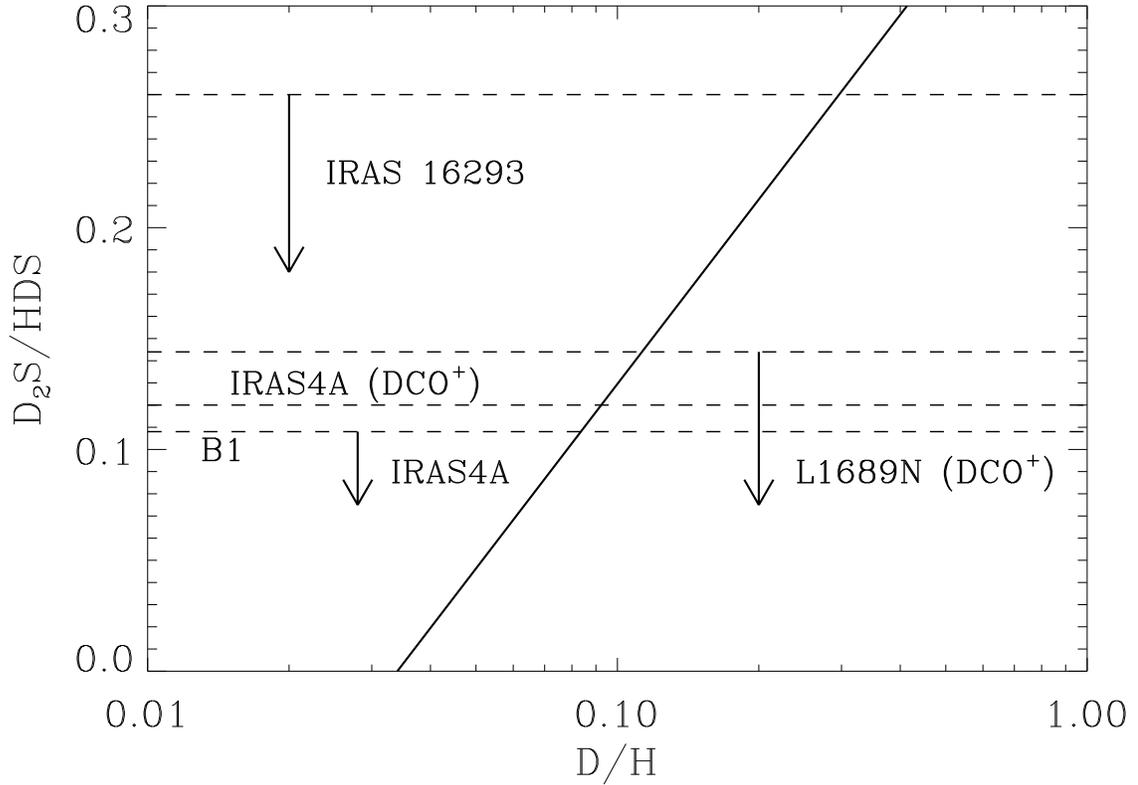}
\caption{D$_2$S/HDS abundance ratio plotted as a function of the D/H abundance (cm$^{-3}$), 
for high density conditions at 10 K using a modified rate approach of formation on 
interstellar grain surfaces \citep[see][]{caselli02}. B1 and NGC1333-IRAS4A are plotted at the 
same D$_2$S/HDS ratio, but B1 is a detection and NGC1333-IRAS4A an upper limit only. 
Upper limits are represented with arrows. \label{fig2}}
\end{figure}

H$_3$S$^+$ is thought to be the precursor to the H$_2$S molecule, via a dissociative 
recombination. The reaction is important because all the 
normal SH$_n$$^+$~+~H$_2$~$\rightarrow$~SH$_{n+1}$$^+$~+~H reactions 
are highly endothermic \citep{millar86} and cannot occur in cold 
interstellar clouds. Nevertheless, even at low temperatures, the 
rate coefficient of H$_3$S$^+$ production is very low 
\citep[$\sim$ 4 $\times$ 10$^{-15}$ cm$^{3}$s$^{-1}$ at 10 K,][]{herbst89}, limiting the 
production of H$_2$S molecules in cold clouds. 
On the contrary, sulfur is expected to easily react with H on grain 
surfaces and thereby form the volatile hydride H$_2$S \citep{tielens87}. 
In the same vein, formaldehyde and methanol are believed to
be formed via active grain chemistry with successive hydrogenations of CO, 
stored on the grain mantles, and eventually released into the gas, for instance 
during the collapse phase, when the heating of newly formed 
protostars evaporates the CO-rich ices.\\
If the deuteration of the three ices (H$_2$S, H$_2$CO and CH$_3$OH) occurs when the ices 
are formed and if it is dominated by pure statistics (i.e. by the random distribution
of D atoms), the deuteration ratios
would assume the following values \citep[detailed discussion in][]{rodgers02}: 
F(H$_2$S) = (HDS/H$_2$S)$^2$/(D$_2$S/H$_2$S) = 4,
F(H$_2$CO) = (HDCO/H$_2$CO)$^2$/(D$_2$CO/H$_2$CO) = 4,
F(CH$_3$OH) = (CH$_2$DOH/CH$_3$OH)$^2$/(CHD$_2$OH/CH$_3$OH) = 3.
Observed values different from these would indicate either different
chemical pathways or chemical reactions on the grain surfaces
which favor D over H (or vice-versa) substitutions.
Note that the modified rate approach and Monte Carlo method by \citet{caselli02} to 
explain deuteration fractionation on interstellar grains indeed
confirm the above values -assuming that the chemistry of S
is similar to that of O on the grain surfaces.\\
Our sample includes IRAS16293-2422, where the singly and doubly
deuterated forms of both formaldehyde and methanol have been observed
\citep{ceccarelli98,loinard00,parise02}.
The observed values are:
F(H$_2$CO) = 0.4 and
F(CH$_3$OH) = 4.
Combining the present D$_2$S/HDS upper limit (Table \ref{table3}) with the
HDS/H$_2$S value found by \citet[HDS/H$_2$S$\sim 0.1$]{vandishoeck95} gives:
F(H$_2$S) $\geq$ 0.5. It seems hence that while methanol and hydrogen sulfide are 
consistent with the hypothesis of the random distribution of D's
and formation on the grain surfaces, formaldehyde is definitively off.
The use of the F factors has the great advantage to highlighting
what chemical mechanism is responsible for the observed deuteration
\citep{rodgers02}, but obviously overlooks the details of it.\\
We computed in Table \ref{table3} the column densities for the HDS molecule. 
Assuming that the HDS/H$_2$S is around 10\%, the value found for formaldehyde 
and other molecules presenting a large degree of deuteration, one would obtain H$_2$S 
abundances in the range (0.2~-~2) $\times$ 10$^{-9}$. This value has been computed with an H$_2$ 
column density about 2.8 $\times$ 10$^{23}$ cm$^{-2}$ for L1689N(DCO$^+$) \citep{lis02}, 
2 $\times$ 10$^{23}$ cm$^{-2}$ for IRAS16293 \citep{vandishoeck95}, 
1.3 $\times$ 10$^{23}$ cm$^{-2}$ for B1 \citep{hirano99} and 
3.1 $\times$ 10$^{23}$ cm$^{-2}$ for IRAS4A \citep{vdt02}. We don't know yet the actual value 
for the H$_2$ column density towards IRAS4A(DCO$^+$). We can say that the computed H$_2$S 
abundance seems 
higher than the range of 10$^{-11}$~-~10$^{-10}$ estimated for steady-state gas-phase models 
\citep{millar90,millar97}, and it would be a strong evidence that grains, and not gas, are 
producing H$_2$S. Of course, this is a rough estimate, and observations for the H$_2$S 
molecule have already been planned in order to derive more accurate values.\\
The diagonal line in Fig. \ref{fig2} shows the predicted D$_2$S/HDS as function of the atomic
D/H ratio in the accreting gas, assuming that D$_2$S/HDS is equal to D$_2$O/HDO,
as computed by \citet{caselli02}. The argument that the surface chemistries of sulfur and 
oxygen are similar can be challenged by the fact that D$_2$O has not yet been detected in 
the interstellar medium. But the search for D$_2$O has not yet been carried out towards the same 
objects where D$_2$S has been detected, and it will be interesting indeed to check whether the 
two molecules, D$_2$S and D$_2$O, have very different abundances with respect to their respective 
main isotopomers. In the model presented by \citet{caselli02}, surface chemistry can convert a 
high atomic D/H ratio in the gas to high abundance ratios for singly, doubly or 
triply deuterated isotopomers, especially at high gas densities (10$^5$ cm$^{-3}$) 
and low temperatures (10 K). 
The observed values in B1 and NGC1333-IRAS4A(DCO+) and upper limits of 
Table \ref{table3} are consistent with previous determinations of the required atomic 
D/H ratio in the accreting gas of $\sim 0.2$~-~0.3 found in the case of IRAS16293-2422 
\citep{parise02}.\\
We conclude that, based on the observations so far available, hydrogen sulfide, formaldehyde 
and methanol are very likely formed on the grain surfaces, and the deuteration occurs during 
the formation of the mantles. What remained problematic to explain up to very recently is the 
relatively large value of the atomic D/H ratio required to achieve the observed large deuteration 
fractionation in the three molecules. In fact until very recently, no gas-phase chemical
model predicted such large values of atomic D/H \citep[e.g.][]{roberts00a,roberts00b}.
However, the current year has seen the detection of an extremely large abundance
of H$_2$D$^+$ in the pre-stellar core L1544 \citep{caselli03}, 10 times larger than prediction 
from previous chemical model. This detection, together with the previous observations
of the doubly and triply deuterated molecules, has stimulated a radical
revision of the chemical gas phase models. \citet{phillips02} pointed out that the 
H$_3^+$ chemistry should include D$_2$H$^+$ and D$_3^+$, as well as the expected H$_2$D$^+$. 
\citet{roberts03} showed that inclusion of the doubly and triply deuterated forms of H$_3^+$ 
would increase the atomic D/H ratio in CO depleted gas to the levels 0.1~-~1, required by the 
D$_2$CO, CHD$_2$OH and the present D$_2$S observations. In conclusion, this work supports the 
hypothesis that formaldehyde, methanol and hydrogen sulfide are indeed formed on the grain 
surfaces, during the cold pre-stellar core phase, where the CO depleted gas possesses large 
atomic D/H ratios.

\clearpage

\clearpage

\begin{deluxetable}{lcc}
\tablecolumns{3}
\tablewidth{0pc}
\tablecaption{Source coordinates.\label{table1}}
\tablehead{
\colhead{Source} & \colhead{RA (2000)} & \colhead{DEC (2000)}} 
\startdata
NGC1333-IRAS2            & 03h28m55.4s   & 31\arcdeg14\arcmin35.1\arcsec\\
NGC1333-IRAS4A (DCO$^+$)\tablenotemark{a} & 03h29m12.1s   & 31\arcdeg13\arcsec26.4\arcsec\\
IRAS16293                & 16h32m22.6s   & -24\arcdeg28\arcsec33.0\arcsec\\
B1                       & 03h33s20.8s   & 31\arcdeg07\arcmin34.4\arcsec\\
NGC1333-IRAS4A           & 03h29m10.3s   & 31\arcdeg13\arcmin32.3\arcsec\\
L1689N (DCO$^+$)\tablenotemark{b}         & 16h32m28.6s   & -24\arcdeg29\arcmin2.7\arcsec\\
\enddata
\tablenotetext{a}{This source corresponds to a position 23$\arcsec$ East and 6$\arcsec$ South 
of the IRAS 4A position, and is where the DCO$^+$ 3 $\rightarrow$ 2 emission peaks (Lis et al., 
in preparation)}
\tablenotetext{b}{This position corresponds to the DCO$^+$ emission peak observed by \cite{lis02} in 
the star forming molecular cloud L1689N}
\end{deluxetable}
\clearpage
\begin{deluxetable}{lcccc}
\tablecolumns{5}
\tablewidth{0pc}
\tablecaption{Line parameters.\label{table2}}
\tablehead{
\colhead{} & \multicolumn{2}{c}{HDS (1$_{0,1}$-0$_{0,0}$)} & \multicolumn{2}{c}{D$_2$S (1$_{1,1}$-0$_{0,0}$)}\\
\colhead{} & \multicolumn{2}{c}{(244 555.58 MHz)} & \multicolumn{2}{c}{(237 903.6752 MHz)}\\
\cline{2-5}\\
\colhead{Object} & \colhead{T$_{mb}$} & \colhead{$\Delta$v} & \colhead{T$_{mb}$} & \colhead{$\Delta$v}\\
\colhead{} & \colhead{(K)} & \colhead{(km/s)} &\colhead{(K)} & \colhead{(km/s)}} 
\startdata
NGC1333-IRAS2            & $<$ 0.05 &     & $<$0.03  &  \\
NGC1333-IRAS4A (DCO$^+$) & 0.31     & 1.4 & 0.08     & 1.8\\
IRAS16293                & 0.13     & 1.4 & $<$ 0.1  & \\
B1                       & 0.22     & 1.1 & 0.07     & 0.6\\ 
NGC1333-IRAS4A           & 0.17     & 1.7 & $<$ 0.05 & \\
L1689N (DCO$^+$)         & 0.15     & 1.5 & $<$ 0.06 & \\
\enddata
\end{deluxetable}
\clearpage
\begin{deluxetable}{lcc}
\tablecolumns{2}
\tablewidth{0pc}
\tablecaption{D$_2$S/HDS ratio and HDS column densities for an excitation temperature between 
10 K and 50 K.\label{table3}}
\tablehead{
\colhead{}   & \colhead{N(D$_2$S)/N(HDS)}   & \colhead{N(HDS)}\\
\cline{2-3}\\
\colhead{Object} & \colhead{($\%$)}         & \colhead{(10$^{13}$ cm$^{-2}$)}}
\startdata
NGC1333-IRAS4A (DCO$^+$) & 12.0       &  1.2~-~4.8\\
IRAS16293                & $<$ 26.4   &  0.5~-~2.0\\
B1                       & 10.8       &  0.7~-~2.6\\
NGC1333-IRAS4A           & $<$ 10.8   &  0.8~-~3.1\\
L1689N (DCO$^+$)         & $<$ 14.4   &  0.6~-~2.5\\
\enddata
\end{deluxetable}

\end{document}